\documentclass[aps,prl,preprint,groupedaddress]{revtex4}

\begin{document}


\title{Possible unconventional superconductivity
in Na$_x$CoO$_2$yH$_2$O probed by $\mu$SR
}


\author{Wataru Higemoto$^1$, Kazuki Ohishi$^1$, Akihiro Koda$^1$,
Ryosuke Kadono$^1$\footnote{Also at School of Mathematical and
Physical Science, The Graduate University for Advanced Studies.}
Kenji Ishida$^2$,
Kazunori Takada$^{3,5}$, Hiroya Sakurai$^4$, Eiji Takayama-Muromachi$^4$
and Takayoshi Sasaki$^{3,5}$}
\affiliation{$^1$Institute of materials structure Science, High Energy
Accelerator Research Organization, Tsukuba, Ibaraki 305-0801,Japan\\
$^2$Department of Physics, Graduate School of Science, Kyoto University,
Kyoto 606-8502, Japan\\
$^3$Advanced Materials Laboratory, National Institute for
Materials Science, Tsukuba, Ibaraki 305-0044, Japan\\
$^4$Superconducting Materials Center, National Institute
for Materials Science,
Tsukuba,Ibaraki 305-0044, Japan\\
$^5$CREST, Japan Science and Technology Agency
}


\date{\today}

\begin{abstract}
The superconducting property of recently discovered sodium cobalt
oxyhydrate, Na$_{0.35}$CoO$_2$1.3H$_2$O ($T_c=4.5$ K),
has been studied by means of muon
spin rotation/relaxation ($\mu$SR) down to 2~K.
It was found that the zero-field muon spin relaxation rate is independent of
the temperature, indicating that no static magnetism appears in this compound,
at least above 2~K. 
The result also provides evidence against the breakdown 
of time-reversal symmetry for the superconducting order parameter. 
Meanwhile, the muon Knight shift at 60~kOe shows no obvious reduction
below $T_c$, suggesting that the local spin susceptibility is  
preserved upon a superconducting transition. 
Considering these observations, possible unconventional
superconductivity of Na$_{0.35}$CoO$_2$1.3H$_2$O is discussed.
\end{abstract}


\maketitle

The recent discovery of superconductivity in a novel
cobalt oxide, Na$_x$CoO$_2$$y$H$_2$O ($x=0.35, y=1.3$),
with the transition temperature $T_c\sim$5~K,
is attracting much interest\cite{Takada}.
The compound has a lamellar structure 
consisting of CoO$_2$, Na and H$_2$O layers,
where the two-dimensional (2D) CoO$_2$ layers are separated
by thick insulating Na or H$_2$O layers.
This structure is similar to
high-$T_c$ cuprate superconductors (HTSCs) in the sense
that they also have a layered structure of
2D-CuO$_2$ sheets separated by
insulating layers.
It is well established that
Cu$^{2+} (S=1/2)$ atoms on a square lattice exhibit antiferromagnetic (AF)
ordering in the parent compounds of HTSCs, where the superconductivity occurs
when the AF state is suppressed by carrier doping.
On the other hand, Co atoms form a 2D
triangular lattice on the CoO$_2$ layers, where a strong magnetic 
frustration is anticipated.
Thus, while Na$_{0.35}$CoO$_2$1.3H$_2$O may be viewed as an electron-doped
Mott insulator for a low-spin Co$^{4+} (S=1/2)$
with electron doping of $x$=35\%, the electronic state may be considerably
different from cuprates.

Although several experiments have revealed interesting properties
of the present system\cite{sakurai,schaak,waki,kobayashi,fujimoto,ishida},
the situation is far from reaching a consensus on the important
issues, including that on the pairing symmetry of superconductivity.
Meanwhile, based on the unique structure of the 2D-triangular Co lattice, 
many theoretical models predicting a variety of
unconventional superconductivity have been proposed.
For example, superconductivity
with symmetries of the $p+ip$ state\cite{Tanaka},
the $d_{x^2-y^2}+id_{xy}$ state\cite{Baskaran, Ogata, Wang, Honerkamp}, 
and the $f$ state\cite{Ikeda} are argued. 
It is notable that some of these states break
the time-reversal symmetry of the Cooper pairs, leading to
the appearance of a weak spontaneous internal magnetic field
in accordance with the superconducting transition.
Such an internal field can be detected with utmost sensitivity
by the zero-field muon spin relaxation (ZF-$\mu$SR) technique.

Furthermore, the muon Knight shift give crucial information about
the pairing symmetry of the Cooper pairs.
Up to now, there are two $^{59}$Co-NMR Knight shift measurements.
Waki {\it et al.} reported a Knight shift that is independent of temperature
through $T_c$, which suggests a spin triplet
state for the pairing symmetry\cite{waki}.
On the other hand, Kobayashi {\it et al.} showed a decrease of the Knight shift
with decreasing temperature\cite{kobayashi}.
The Knight shift results are presently a controversial issue.
Thus, it is quite important to obtain more clues
brought by other experimental techniques, including $\mu$SR,
to settle the issue of paring symmetry.
Since the muon has a spin of 1/2, one can
deduce the muon Knight shift without any
complication due to electric field gradient. This feature
is an advantage of the
muon Knight shift over NMR at $^{59}$Co ($I=7/2$) in which
a complex signal pattern is observed.

In this letter, we report on the magnetic and superconducting
properties of Na$_{0.35}$CoO$_2$1.3H$_2$O studied by means of
the muon spin rotation/relaxation method ($\mu$SR) down to 2~K.
It is inferred from the ZF-$\mu$SR measurement that there is 
no appreciable static magnetism over the entire temperature range
across $T_c$, indicating that the time-reversal symmetry is
preserved in the superconducting state. The muon Knight shift at
60 kOe is almost independent of the temperature irrespective of the superconducting
transition, which suggests
a spin triplet symmetry 
for the order parameter.

Powder specimens of Na$_{0.35}$CoO$_2$1.3H$_2$O,
including a deuterated one (D$_2$O$\simeq75$ \%), were
synthesized, as described in Ref.\cite{Takada}. 
Each specimen was characterized by measuring the magnetic susceptibility 
prior to a $\mu$SR measurement.
Conventional $\mu$SR measurements under zero field and a weak transverse field 
(TF, $H=374$~Oe) were carried out at the $\pi$A-port of the Meson Science
Laboratory, High Energy Accelerator Research Organization (KEK). 
The muon Knight shift measurements (TF-$\mu$SR at 60~kOe) were performed on the
M15 beamline of TRIUMF.  In both cases, 
a positive muon beam with a momentum of 27 MeV/c was implanted to a
powder specimen placed in a He-gas exchange cryostat, where
special precaution was taken to cool down the specimen rapidly below
$\sim$100 K to preserve its water content.
For ZF-$\mu$SR, residual field was reduced to below 10~mOe by
using three pairs of correction magnets.
For a high-field measurement at 60~kOe,
a powder specimen was secured with Apiezon-N grease to prevent
an alignment of fine crystals in the specimen by a strong field.

Fig. 1 shows the time evolution of the muon spin polarization in
Na$_{0.35}$CoO$_2$1.3H$_2$O at 10~K and 2~K
under zero magnetic field.
At 10~K, the muon spin depolarizes due to
a static random local field, which
originates from the  $^{59}$Co, $^{23}$Na and $^1$H nuclear 
magnetic moments.
These spectra can be described by the
Kubo-Toyabe relaxation function, $G_{KT}(\Delta, \nu, t)$\cite{Hayano},
as indicated by the solid line in Fig.1.
Here, $\Delta/\gamma_{\mu}$ is the second moment 
of the field distribution at the muon site, with $\gamma_{\mu}$
being the muon gyromagnetic ratio (=2$\pi\times 13.55$~kHz/Oe),
and $\nu$ is a fluctuation rate of the nuclear dipolar field.
From a fitting analysis, we obtained $\Delta/\gamma_{\mu}\sim$5.0~G
and $\nu\sim 0.22~\mu$s$^{-1}$ at 10~K.
By comparing the experimental
value of $\Delta$ in the normal phase (10~K) of
Na$_{0.35}$CoO$_2$1.3H$_2$O($\Delta=0.425(4)$ $\mu$s$^{-1}$) and
that in the deuterated specimen ($\Delta=0.243(3)$ $\mu$s$^{-1}$, also appears in Fig.1),
together with calculated mapping of $\Delta$, 
the muon stopping site was identified near (0.2,0.25,0.12).

As evident in Fig.~1, we observed no significant change in the 
ZF-$\mu$SR time spectrum while the temperature passed $T_c$.
Fig.~2 shows the temperature dependence of the
dipolar width ($\Delta$), which is nearly independent
of the temperature within an accuracy of 0.1~Oe.
This result clearly demonstrates
the absence of static magnetism over the time window of $\mu$SR
(10$^{-9}\sim 10^{-5}s$).
The upper bound
for the possible magnetic moment was estimated 
to be 0.001$\mu_B$/Co in
Na$_{0.35}$CoO$_2$1.3H$_2$O
by using the hyperfine coupling constant,
$A_{hf}=-134$~Oe/$\mu_B$, which we derive
later.

It is theoretically suggested that spontaneous magnetic fields
are induced below the superconducting transition when the superconductivity
is carried by the `magnetic' Cooper pairs. 
For example, Wang {\it et al.} proposed
that the system falls into a pairing state described by
$d_{x^2-y^2}+id_{xy}$, and that the orbital
current produces a spontaneous magnetic field, which is estimated
to be 10-100~Oe\cite{Wang}. Tanaka and Hu proposed
a chiral $p$-wave state, which also breaks the time-reversal
symmetry \cite{Tanaka}.
Such a spontaneous field has been
observed in a spin-triplet superconductor, Sr$_2$RuO$_4$\cite{Luke1,
HigemotoSRO}, and quite recently in a heavy fermion superconductor, PrOs$_4$Sb$_{12}
$\cite{Aoki}.
However, the results given in Figs.~1 and 2 indicate that there
is no such enhancement of $\Delta$ in the ZF-$\mu$SR
spectra below $T_c$ in Na$_{0.35}$CoO$_2$1.3H$_2$O,
which demonstrates the absence of an additional
spontaneous magnetic field above 0.1~Oe in the superconducting phase.
This result strongly disfavors a pairing symmetry,
like $d_{x^2-y^2}+id_{xy}$ in Na$_{0.35}$CoO$_2$1.3H$_2$O.

Prior to the muon Knight shift measurement, 
we performed  weak transverse field $\mu$SR measurements
to evaluate the effect of the flux line lattice (FLL) in the mixed state.
At 374~Oe, the dumping rate of the spin precession signal
slightly increases with decreasing temperature below
$T_c$, which reflects the formation of FLL, as follows.
In the mixed state of a type-II
superconductor, an applied magnetic field ($H_{c1}<H<H_{c2}$,
with $H_{c1}$ and $H_{c2}$ being the lower and upper critical
field, respectively) induces FLL, where
the internal magnetic field distribution is determined by the
magnetic penetration depth ($\lambda$), the
vortex core radius, and the lattice structure of FLL.
When the system falls into the
superconducting state with
a typical length scale of $\lambda$ from 10$^2$ to 10$^3$ nm,
it leads to additional dumping of the muon spin
precession due to the inhomogeneity of the local field associated
with the FLL state.  Unfortunately, 
$\lambda$ in Na$_{0.35}$CoO$_2$1.3H$_2$O
turned out to be too long to deduce detailed information about
the structure of vortices, such as the core radius.
In such a situation, one can represent the muon spin relaxation
approximately by a simple Gaussian relaxation, namely
\begin{eqnarray}
G_x(t) &=&A_s\left(f_s\exp\left(-\sigma_{FLL}^2t^2\right)
+\left(1-f_s\right)\right)
\exp\left(-\sigma_s^2t^2\right)\cos\left(\omega_st+\phi\right)\nonumber\\ 
& &+A_b\exp\left(-\Lambda_bt\right)\cos\left(\omega_b t+\phi\right),
\end{eqnarray}
where the first term corresponds to the signal
from the specimen and the second term
to that from the backing silver. The damping factor,
$\sigma_s$(=0.21(1)$\mu s^{-1}$ at 10~K),
comes from the relaxation in the normal
phase, which is determined by random nuclear dipolar fields, and is
thus proportional to $\Delta$, measured by ZF-$\mu$SR
(i.e., $\Delta=\sqrt5\sigma_s$\cite{Hayano}).
$f_s$ denotes the volume fraction of the superconducting part
in the present specimen, which is estimated to be
$\sim$0.5 at 2.0K from the fitting result.
Note that
the relaxation rate of backing silver ($\Lambda_b$) is negligibly small.
The internal field distribution probed by muons is a convolution
of the nuclear dipolar field distribution, and that
due to FLL and $\sigma_{FLL}$ indicate the effect of FLL.
Fig. 3 shows the temperature dependence of
$\sigma_{FLL}$ at 374~Oe, where $\sigma_{FLL}$ increases 
with decreasing temperature below $T_c$.
In an isotropic superconductor with a hexagonal FLL,
the second moment ($\langle\Delta B^2\rangle$) is approximately
given by \cite{Brandt1988}
\begin{equation}
 \langle\Delta B^2\rangle=2\sigma_{FLL}^2/\gamma_{\mu}\approx7.5\times10^{-4}
(1-h)^2(1+3.9(1-h)^2)\Phi_0^2\lambda^{-4},
\end{equation}
where $h=H/H_{c2}$ and $\Phi_0$ is the magnetic flux quantum.
From these relations, the average
penetration depth ($\lambda$) is estimated to be
$7(1)\times10^2$~nm
at 2~K under $H=374$~Oe.

The presence of line nodes on the superconducting
gap is suggested from recent Co-NQR measurements\cite{fujimoto, ishida}.
Such a structure of the order parameter should
be reflected on the temperature/field dependence of the
penetration depth.
Unfortunately, the observed $\lambda$ in this compound is too long to
obtain such information with reliable sensitivity.

The field inhomogeneity $\langle \Delta B^2\rangle$
decreases with increasing field
due to the strong overlap of the flux lines and the increased contribution
of vortex cores,
leading to a diminishingly small spin relaxation ($\sigma_{FLL}\sim 0$)  
at higher fields.  This tendency is more enhanced when the
superconducting order parameters has a nodal structure.
In this situation, we can not distinguish the superconducting fraction from
the normal fraction in the specimen, and thereby
the spectra at 60~kOe can be analyzed by using the following simple relation:
\begin{equation}
G_x(t)=A_s
\exp(-\sigma_s^2t^2)\cos(\omega_st+\phi),
\end{equation}
where we neglect the signal from the small amount of Apiezon-N grease,
and omit the background term ($A_b$) because the measurements
at 60 kOe are made without silver backing.
The relaxation rate ($\sigma_s$), where the
obtained data are shown in Fig.~3
by open circles, is independent of the temperature, indicating that 
the effect of FLL is negligible.
In this case, the frequency shift due to FLL is also estimated to be negligibly small
(see below) and the muon Knight shift can be observed.
Since $H_{c2}$ is reported to be 610~kOe with
$T_c$ staying almost constant up to 40~kOe by a magnetization
measurement\cite{sakurai}, 
the system
should
be in the superconducting state 
at 60~kOe.
In general, the muon Knight shift $K(T)$ is expressed as
$K(T)=K_s(T)+K_{orb}$. Here, $K_s$ and $K_{orb}$ are
the spin and orbital components of the Knight shift
and only $K_s(T)$ is temperature dependent.
Then, $K_s(T)$ is determined by the gradient  of
the muon Knight shift versus the magnetic susceptibility ($K$-$\chi$) plot,
as shown in Fig.~4(a), irrespective of $K_{orb}$
(which was set to zero in this analysis).
Fig. 4(b) shows the temperature dependence of $K_s(T)$ 
and the uniform susceptibility at 10~kOe in a
randomly oriented sample of Na$_{0.35}$CoO$_2$1.3H$_2$O.
The muon Knight shift decrease with decreasing temperature below 100~K,
and levels off below 10K.
Above 10~K, $K_s(T)$ is proportional to the
uniform susceptibility, which clearly indicates that the 
upturn of the $\chi(T)$ below
$\sim100$~K is not due to impurities,
but due to some intrinsic origin.
The spin part of the muon Knight shift is expressed as
$K_s=A_{hf}\chi$,
where $A_{hf}$ is the hyperfine coupling constant.
From the above relation,
$A_{hf}$ is estimated to be $\sim-134$~Oe/$\mu_B$.
It should be noted that the anisotropic term of the Knight
shift, or the so-called powder pattern, is not seen in the spectra.
This implies that the anisotropy of the Knight shift is too small
for the present resolution.
In the case of $s$-wave pairing superconductivity,
$K_s$ decreases to zero with decreasing temperature
following the Yosida function\cite{Yoshida}.
However, as shown in the inset of Fig. 4,
the muon Knight shift does not show any appreciable reduction
below $T_c$.
Indeed, the average of $K_s$ for 4.7-10~K ($>T_c$)  is $-14.4(5)$~ppm
which is in good agreement with $ -15.1(5)$~ppm for 2.0-4.4 K ($<T_c$).
This result 
shows that the spin susceptobility is unchanged in the superconducting
state.

Since the observed muon Knight shift is negative, there remains a
possibility that the reduction of $K_s$ below $T_c$
may be cancelled by the diamagnetic shift upon the
formation of FLL.
We estimated the magnitude of such a diamagnetic
shift by taking account of the effects of a dense overlap of vortices
and the Doppler shift; the latter is present when the superconducting
gap has nodes, leading to a further enhancement of $\lambda$\cite{Volovik}.
This is reasonable considering the recent NQR
measurements\cite{fujimoto, ishida}, suggesting line nodes
in this compound.
Our simulation yielded that $\lambda\simeq12$ $\mu$m,
where the associated diamagnetic shift by the FLL formation
is less than 2.8~ppm at 60~kOe, which
is too small to cancel out the  predicted effect due to spin 
singlet superconductivity.  Thus, we conclude that the
result given in Fig.~4 is entirely attributed to $K_s$.

Here, we discuss the symmetry of the Cooper pair.
We observed no significant reduction of the
spin part of the muon Knight shift within the error bar below 10~K.
If we assume that the residual density of state
is 35\%, as estimated from one of the reported 
NMR measurements\cite{ishida},
the reduction in the Knight shift is too small
for the spin singlet superconductivity.
This suggests that the paring of the Cooper pair is a spin triplet state,
which is in line with the result by Waki {\it et al.}, who 
observed no reduction of the $^{59}$Co Knight
shift below $T_c$.
Quite recently, a nearly ferromagnetic spin fluctuation 
was proposed by Ishida {\it et al.}
These observations seem to be promising for spin triplet
superconductivity from the analogy of the superfluidity
of $^3$He.
Our result is inconsistent with the Knight-shift results in Ref [4],
which was measured in 
15~kOe. The inconsistency might be related to the difference
of the applied field.

Theoretically, $p+ip$ or $f$ wave superconductivity is
suggested for the spin triplet superconductivity.
Since our result of ZF-$\mu$SR excludes 
the possibility of time-reversal
symmetry breaking superconductivity, the chiral
$p$-wave superconductivity is strongly disfavored.
When the spin triplet $d$-vector is pinned to some
direction, it is expected that 1/3($d // z$-axis) or 2/3 ($d\perp z$)
of the muon Knight shift would decrease in a polycrystalline
specimen. The absence of
such a reduction in the Knight shift 
suggests that the effective spin-orbit
coupling is much weaker than 60~kOe, leading
to an alignment of the $d$ vector against the
external field.

In summary, we have demonstrated  by ZF-$\mu$SR that 
no static magnetism appears in  Na$_{0.35}$CoO$_2$1.3H$_2$O
down to 2K. This leads us to conclude that the time-reversal
symmetry is preserved in the superconducting order parameter. 
The muon Knight shift at 60~kOe exhibits no obvious reduction
below $T_c$ down to 2~K, suggesting that the spin part of the
Cooper pair is a spin triplet state.

We thank the staff of KEK-MSL and TRIUMF for technical support
during experiment. This work was partially supported by a Grand-in-Aid
for Scientific Research on Priority Areas, Ministry of Education, Culture,
Sports, Science and Technology, Japan.

\newpage

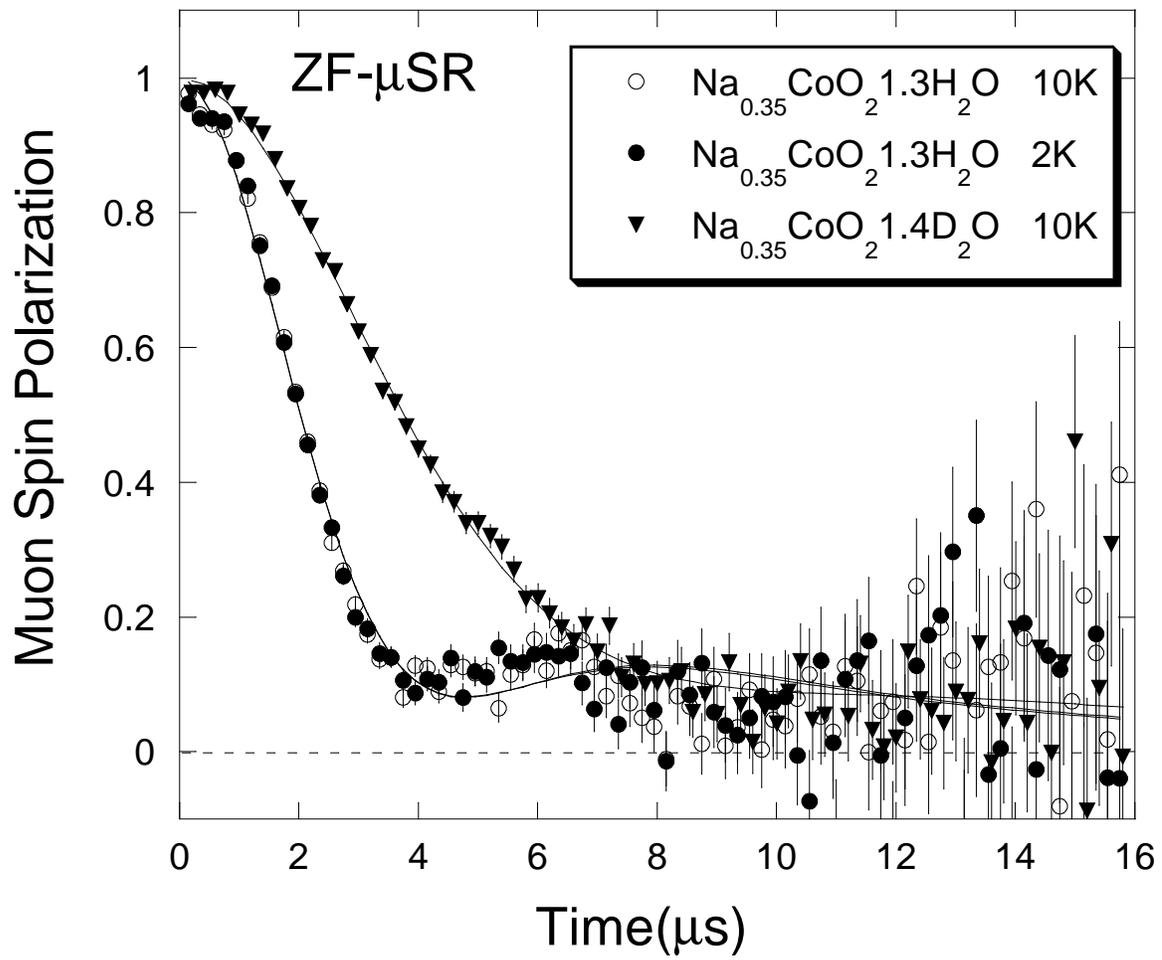
\begin{figure}
\caption{\label{fig1}ZF-$\mu$SR time spectra in Na$_{0.35}$CoO$_2$1.3H$_2$O,
above and below $T_c$.  A spectrum observed at 10~K in a deuterated specimen
is also shown for a comparison.}
\end{figure}

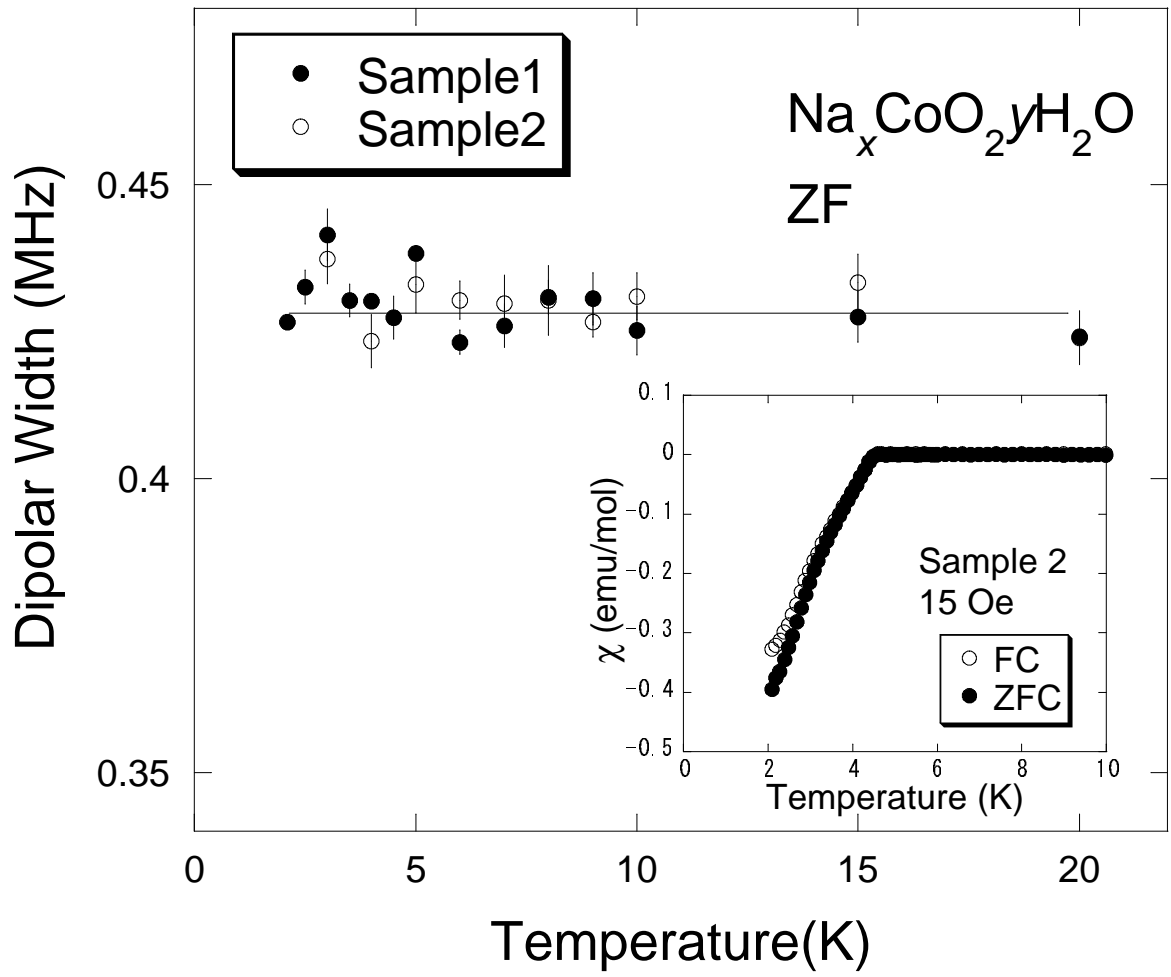
\begin{figure}
\caption{\label{fig2}Temperature dependence of the nuclear dipolar width
($\Delta$) for two specimens from different batches.
Inset: temperature dependence of magnetic
susceptibility with field cooling(FC) and zero field cooling
(ZFC) conditions for the sample \#2.}
\end{figure}

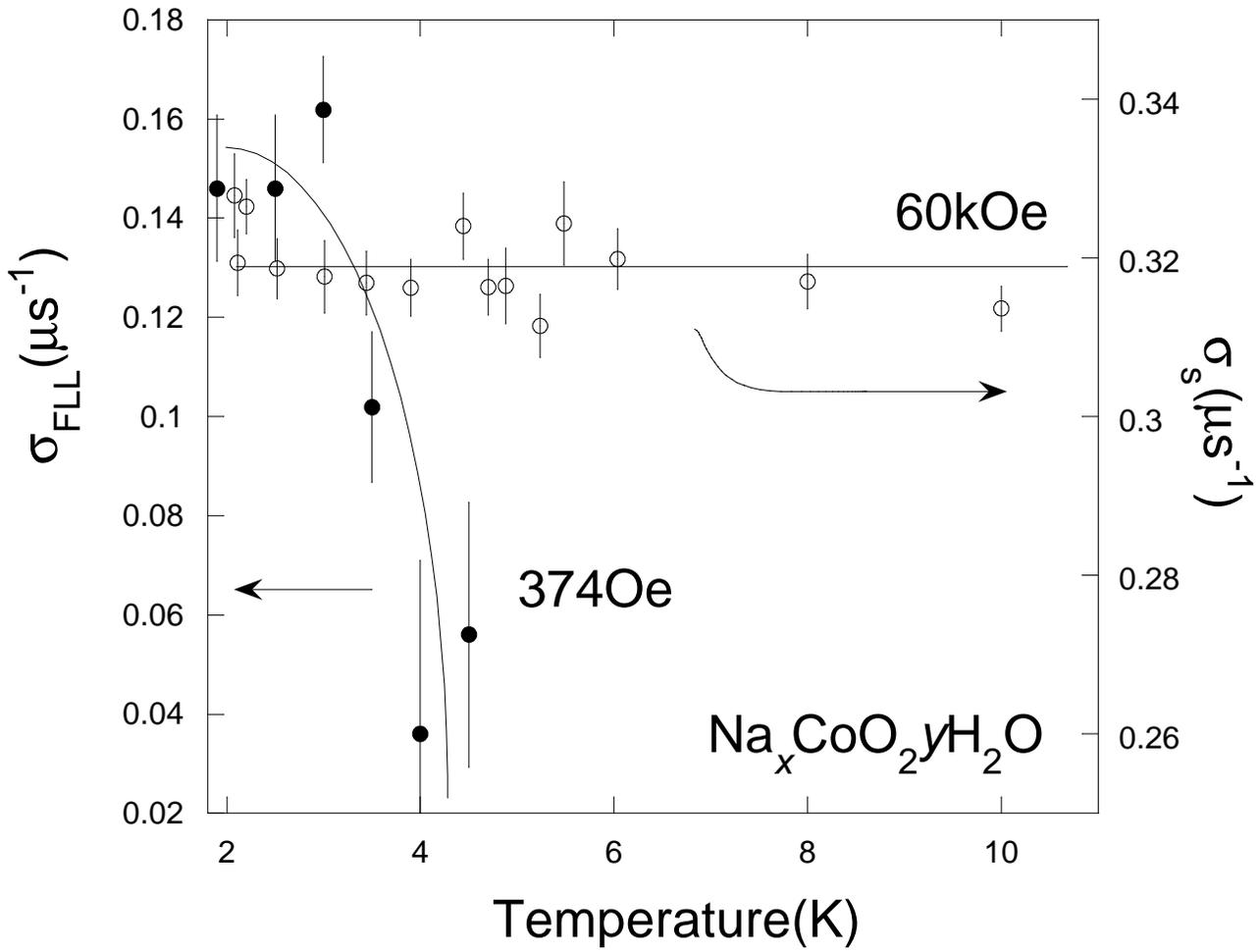
\begin{figure}
\caption{\label{fig3}
Temperature dependence of $\sigma_{FLL}$ at 374~Oe and 
$\sigma_s$ at 60~kOe in Na$_{0.35}$CoO$_2$1.3H$_2$O. 
Solid lines are guide for eyes.}
\end{figure}

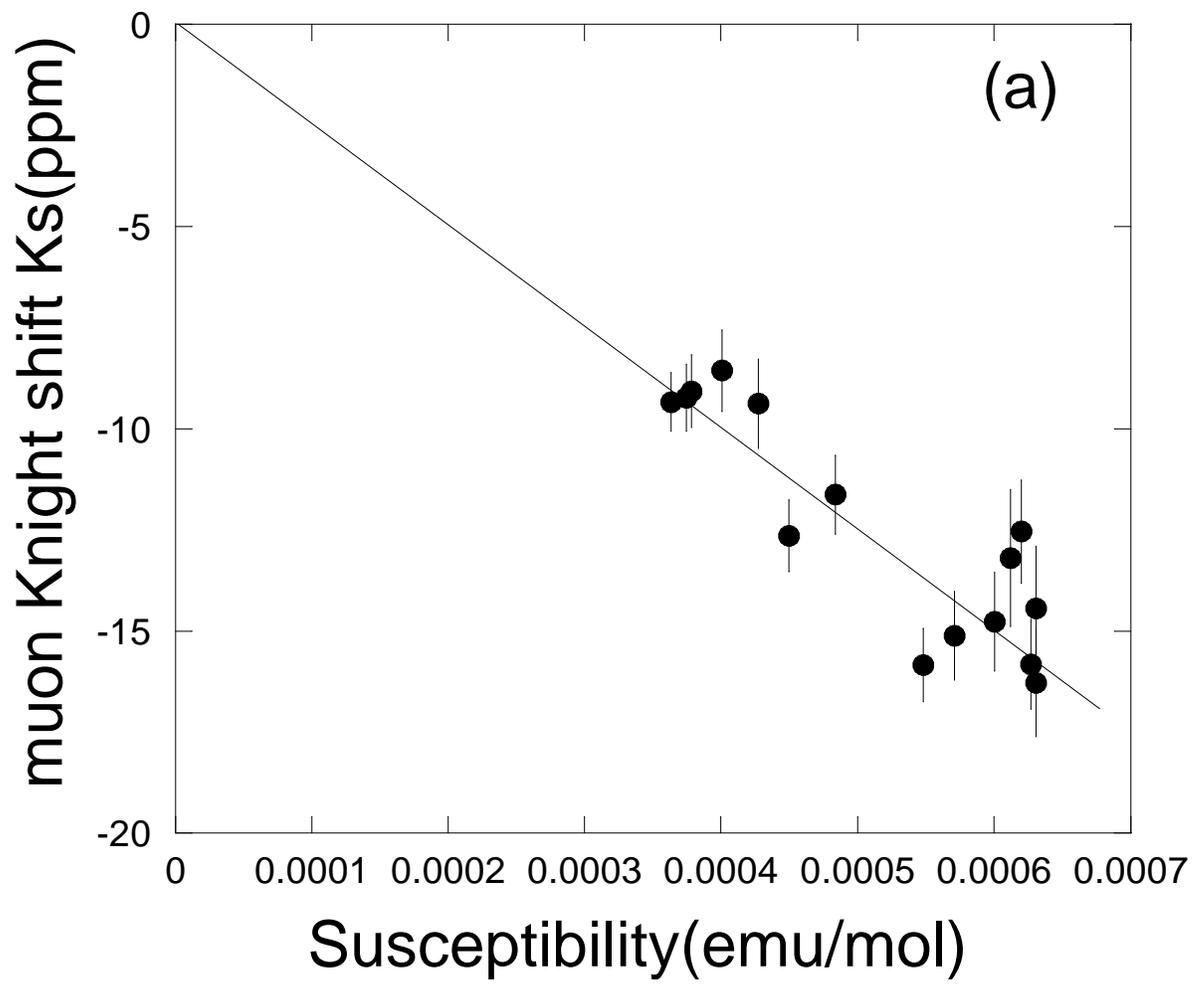
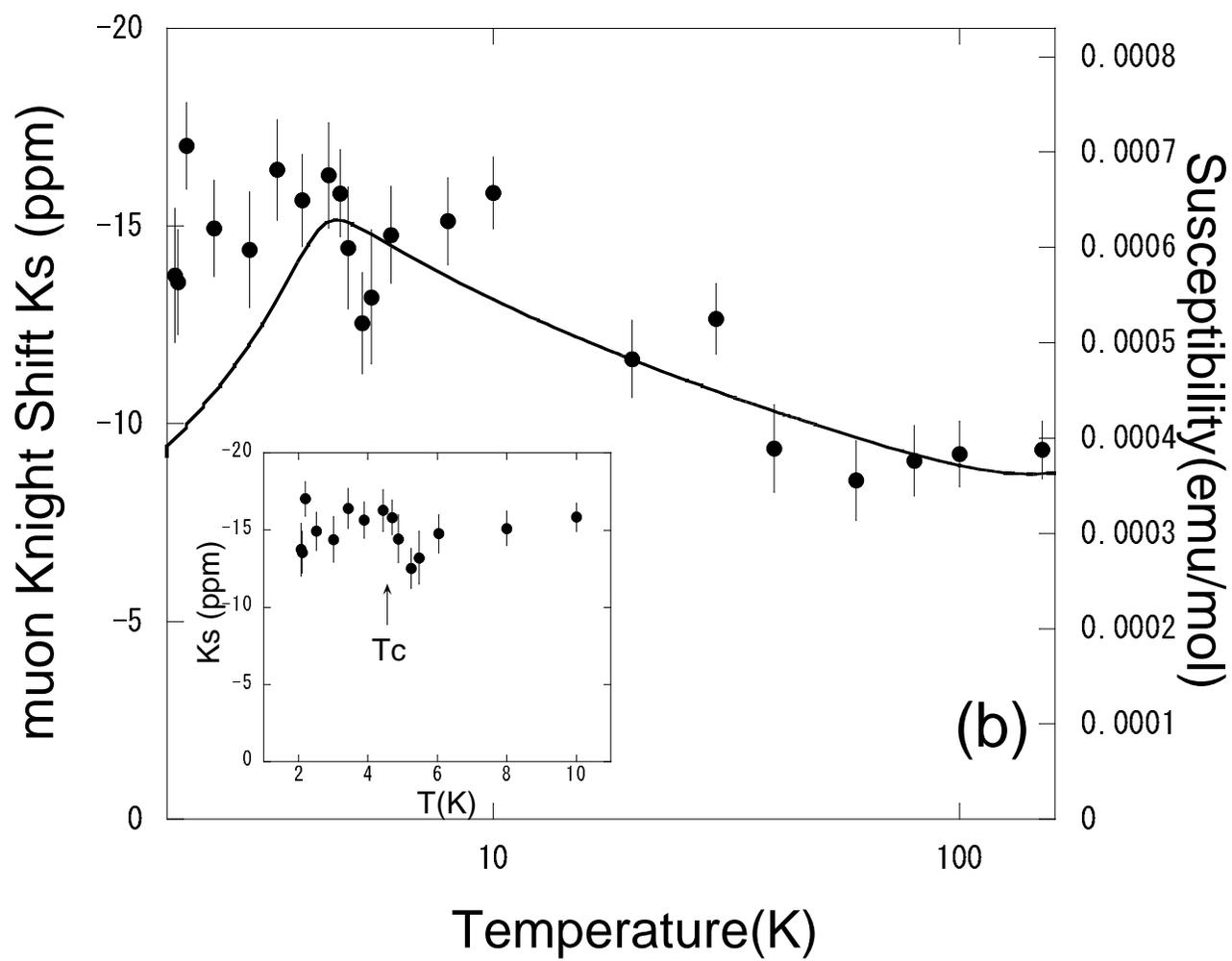
\begin{figure}
\caption{\label{fig4}(a) Muon Knight shift versus susceptibility plot.
(b)Temperature dependence of the spin part of the muon Knight shift $K_s$ (filled dot)
and susceptibility(solid line). The susceptibility was measured at 10~kOe.}
\end{figure}



\end{document}